\documentclass[epj,referee]{svjour}
\usepackage{graphicx}
\usepackage{dcolumn}
\usepackage{bm}
\usepackage{amsmath}
\begin{document}
\title{Reexamination of Astrophysical Resonance-Reaction-Rate Equations for An Isolated, Narrow Resonance}
\author{J.J. He
\thanks{\emph{Email: jianjunhe@impcas.ac.cn} }
\and J. Hu
\and S.W. Xu
\and X.Q. Yu
\and L. Li
\and L.Y. Zhang
}                     

\institute{Institute of Modern Physics, CAS, Nanchang Road 509, Lanzhou 730000, China}

\titlerunning{Reexamination of Astrophysical Resonance-Reaction-Rate ...}
\authorrunning{J.J. He et al.}

\date{Received: date / Revised version: date}
%
\abstract{
The well-known astrophysical resonant-reaction-rate ({\tt RRR}) equations for an isolated narrow resonance induced by the charged particles have been reexamined.
The validity of those `look reliable' assumptions used in deriving the classical analytic equations has been checked, and found that these analytic equations only
hold for certain circumstances. It shows the customary definition of `narrow' is inappropriate or ambiguous in some sense, and it awakes us not to use those analytic
equations without caution. As a suggestion, it's better to use the broad-resonance equation to calculate the {\tt RRR} numerically even for a narrow resonance.
The present conclusion may influence some work in which the classical narrow-resonant equations were used for calculating the {\tt RRR}s, especially at low stellar temperatures for those previously defined `narrow' resonances.
\PACS{
      {25.40.Ny}{Resonance reactions}   \and
      {26.}{Nuclear astrophysics}
     } 
} 
\maketitle

\section{Introduction}
Nuclear astrophysics addresses some of the most remarkable questions in nature: What are the origins of the elements making life on earth possible? How did the sun,
the solar system, and the stars form and evolve? and {\em etc.}. In order to answer these questions, a huge amount of nuclear physics information is required in the
astrophysical models as input. Thereinto the reaction rate, which is defined as how fast a reaction takes place, is one of the most critical ones. It reflects most
of the critical stellar features (such as time scale, energy production as well as nucleosynthesis of the elements, {\it etc\/}). Nuclear astrophysics has been
developed for more than 50 years, a great number of stellar reaction rates have been calculated by utilizing the classical equations based on the experimental data.
However, few people has questioned the validity of those classical reaction-rate equations derived in the historic books or references,
\textit{e.g.}, in~\cite{bib:fow67,bib:cla83,bib:rol88}.

This work is focusing on the validity of resonant-reaction-rate (hereafter referred to as {\tt RRR}) equations for an isolated narrow resonance induced by the charged
particles. Actually these equations have been derived in many books or references. As an example, let's thumb through the famous Book
``{\it Cauldrons in the Cosmos\/}" ~\cite{bib:rol88}. To make readers easy to follow, a similar deriving procedure and the same notations used in~\cite{bib:rol88}
are being adopted here.

It is well-known the stellar reaction rate per particle pair $\langle \sigma v \rangle$ can be written as (Equ.~(4.1) in ~\cite{bib:rol88})
\begin{equation}
\langle \sigma v \rangle = \left (\frac{8}{\pi \mu}\right)^{1/2}\frac{1}{(kT)^{3/2}}\int^{\infty}_{0}\sigma(E)E \mathrm{exp} \left(-\frac{E}{kT} \right)dE.
\label{eq:1}
\end{equation}
As for an isolated resonance the cross section can be expressed by a famous Breit-Wigner formula as (Equ.~(4.52) in \cite{bib:rol88})
\begin{eqnarray}
\sigma(E)_{BW}^N = &&\pi (\frac{\lambda}{2\pi})^2 \frac{2J+1}{(2J_1+1)(2J_2+1)} (1+\delta_{12}) \nonumber \\
&& \times \frac{\Gamma_a \Gamma_b}{(E-E_R)^2+(\Gamma/2)^2}.
\label{eq:2}
\end{eqnarray}
With the knowledge of the energy dependence of the cross section for a narrow resonance ($\Gamma$$\ll$$E_R$), Equ.~\ref{eq:1} of the stellar {\tt RRR} per
particle pair can be replaced by
\begin{equation}
\langle \sigma v \rangle = \left (\frac{8}{\pi \mu}\right)^{1/2}\frac{1}{(kT)^{3/2}}\int^{\infty}_{0}\sigma(E)_{BW}^N E \mathrm{exp} \left(-\frac{E}{kT} \right)dE.
\label{eq:3}
\end{equation}
Assuming the Maxwell-Boltzmann function $E\mathrm{exp}(-E/kT)$ changing very little over the resonance region (hereafter referred to as ``the FIRST assumption"),
Equ.~\ref{eq:3} can be written in a form of, {\it i.e.\/}, Equ.~(4.53) in~\cite{bib:rol88},
\begin{equation}
\langle \sigma v \rangle = \left (\frac{8}{\pi \mu}\right)^{1/2}\frac{1}{(kT)^{3/2}}E_R \mathrm{exp} \left(-\frac{E_R}{kT}\right)\int^{\infty}_{0}\sigma(E)_{BW}^N dE.
\label{eq:4}
\end{equation}
And then the integration of the Breit-Wigner cross section yields, for a narrow resonance with $(\frac{\lambda}{2\pi})^2 \simeq (\frac{\lambda}{2\pi})_R^2$
(hereafter referred to as ``the SECOND assumption") and with negligible energy dependence of the partial and total widths (hereafter referred to as ``the THIRD
assumption"), {\em i.e.}, Equ.~(4.54) in~\cite{bib:rol88},
\begin{eqnarray}
\int^{\infty}_{0}\sigma(E)_{BW}^N dE & = & \pi (\frac{\lambda}{2\pi})_R^2 \omega\Gamma_a\Gamma_b\int^{\infty}_{0} \frac{1}{(E-E_R)^2+(\Gamma/2)^2} \nonumber \\
                                     & = & 2\pi^2 (\frac{\lambda}{2\pi})_R^2 \omega \frac{\Gamma_a\Gamma_b}{\Gamma}.
\label{eq:5}
\end{eqnarray}
Where $\omega\gamma$=$\omega\frac{\Gamma_a\Gamma_b}{\Gamma}$ is defined as the resonance strength.
Finally, the {\tt RRR} per particle pair for an isolated narrow resonance can be written as
\begin{equation}
\langle \sigma v \rangle = \left (\frac{2\pi}{\mu kT}\right)^{3/2} \hbar^2 (\omega \gamma)_R \mathrm{exp} \left(-\frac{E_R}{kT}\right).
\label{eq:6}
\end{equation}
This classical analytic Eq.~\ref{eq:6} is so famous that it has been utilized in a great number of nuclear astrophysics work.

After a careful survey, we find the definition and border of narrow and broad resonances in the previous literature are not quite clear. For instance, a resonance with
$\Gamma/E_R$$\geq$10 \% called a broad resonance~\cite{bib:rol88} implies that a resonance with $\Gamma/E_R$$<$10 \% can be treated as a narrow one, is it appropriate?
In the present work, we will examine the validity of all three assumptions mentioned above.

\begin{figure}[t]
\begin{center}
\includegraphics{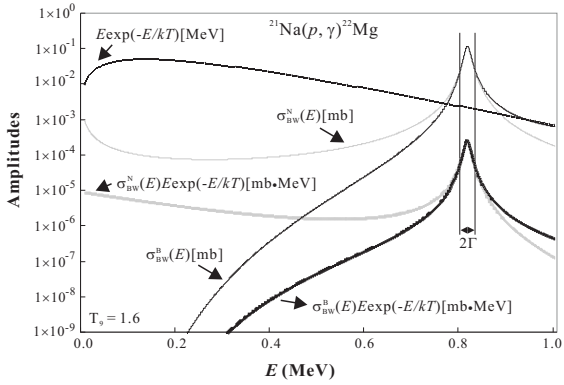}
\vspace{-3mm}
\caption{\label{fig1} Amplitudes of relevant functions used in the equations. The units are included in the square brackets. See text for details.}
\end{center}
\end{figure}

\section{Validity of Assumptions}
\paragraph{The FIRST assumption}
Chronically a resonance for which $\Gamma/E_R$$<$10 \% was treated as a narrow resonance ({\em e.g.}, see ~\cite{bib:rol88}), here it shows this definition of `narrow' is inappropriate
especially at low temperatures. If one defines the variation $\delta_{MB}$ of the Maxwell-Boltzmann function $E\mathrm{exp}(-E/kT)$ over the resonance region as
\begin{eqnarray}
\delta_{MB} &=& 1-\frac{E_R+\Gamma/2}{E_R}\mathrm{exp}\left( -\frac{E_R+\Gamma/2-E_R}{kT} \right) \nonumber \\
            &=& 1-\left(1+\frac{\Gamma/2}{E_R}\right)\mathrm{exp}\left(-\frac{\Gamma}{2kT} \right), \nonumber
\end{eqnarray}
then $\delta_{MB}$ may be large at certain circumstances ({\it e.g.\/}, assuming a resonance with $\Gamma$=16 keV and $\frac{\Gamma}{E_R}$=10 \%,
$\delta_{MB}$$\approx$59 \%, 17 \% at $T$ = 0.1, 0.4 GK, respectively). Clearly, the variation $\delta_{MB}$ exponentially depends on the stellar temperature, and
therefore this assumption is not a good one for low temperatures. From the definition of $\delta_{MB}$, another complementary restriction for a narrow resonance
should be expressed as $\Gamma \leq \frac{T_9}{11.608}\ln\frac{1+\frac{\Gamma/2}{E_R}}{1-\delta_{MB}}$ ($\Gamma$ in unit of MeV). For instance, with
$\delta_{MB}$=5 \% and $\Gamma/E_R$$<$10 \%, another restriction for width is $\Gamma \leq$ 0.017 $T_9$ MeV.
It should be noted that in some literature a definition of `narrow' was $\Gamma$$\ll$$E_R$, but its ambiguity may give rise to certain misleading in some sense.

\paragraph{The SECOND assumption}
The de Broglie wavelength squared $(\frac{\lambda}{2\pi})^2$ (=$\frac{\hbar^2}{2\mu E}$) is inversely proportional to the center-of-mass energy $E$, and therefore this
assumption $(\frac{\lambda}{2\pi})^2$$\simeq$$(\frac{\lambda}{2\pi})_R^2$ means $E$$\simeq$$E_R$ over the resonance region. It's a reasonable assumption for a
resonance with $\Gamma/E_R$$<$10 \%.

\paragraph{The THIRD assumption}
The partial width for a charged particle $a$ (usually either proton or $\alpha$-particle) is strongly energy dependent because the projectile must penetrate the
Coulomb barrier, and can be defined as (Equ.~(4.65) in~\cite{bib:rol88})
\begin{eqnarray}
\Gamma_a=\frac{2\hbar}{R_n} \left(\frac{2E}{u}\right)^{\frac{1}{2}} P_{\ell} \theta^2_\ell, \nonumber
\end{eqnarray}
where the penetration factor can be calculated by $P_{\ell}$=$\frac{1}{(F_{\ell}^2+G_{\ell}^2)}$~\cite{bib:bar74} with the regular and irregular Coulomb wave functions
$F_{\ell}$, $G_{\ell}$. The quantity $\theta^2_\ell$, called the dimensionless reduced width of a nuclear state, is generally determined experimentally and contains
the nuclear structure information. Generally speaking, $\theta^2_\ell$ is independent of energy $E$. Here, for simplicity, only $s$-wave case has been discussed.
Therefore, it is sufficient to use the expression $\Gamma_a$=$c_1 \sqrt{E} P_{\ell=0}(E)$$\propto$$\sqrt{E} e^{-2\pi \eta}$, where the penetration factor $P_{\ell=0}$
for the Coulomb barrier is approximated by a Gamow factor~\cite{bib:rol88}. The quantity $c_1$ depends on the nuclear structure of the resonance. The exponent is
$2\pi \eta$=$c_2\sqrt{1/E}$ with $c_2$=31.29$Z_1 Z_2 \sqrt{\mu}$ depending on the projectile-target combination. Note only in this equation
the center-of-mass energy $E$ is given in units of keV.
On the other hand, $\gamma$-ray partial widths depend only weakly on the energy via $\Gamma_\gamma$=$c_3 E_\gamma^{2L+1}$$\propto$$(E+Q)^{2L+1}$, where $E_\gamma$ and
$L$ are the energy and multipolarity, respectively, of the $\gamma$-ray transition under consideration, and $Q$ is the reaction $Q$-value. The quantity $c_3$ depends
on the nuclear structure of the two levels involved in the interaction. Therefore, for the usual particle-gamma reaction channel ({\em {e.g.}}, $(p,\gamma)$ or
$(\alpha, \gamma)$), the variation of quantity $\Gamma_a\Gamma_\gamma$ ($\propto$$\sqrt{E} (E+Q)^{2L+1} e^{-c_2 \sqrt{1/E}}$) over the resonance region can be
expressed as
\begin{eqnarray}
\delta_{\Gamma_a\Gamma_\gamma} =&& \sqrt{1+\Delta E/E_R} \times \left( 1+\frac{\Delta E}{E_R + Q} \right)^{2L+1} \nonumber \\
&& \times \mathrm{exp} ^{-\frac{c_2}{\sqrt{E_R}}\left(\frac{1}{\sqrt{1+\Delta E/E_R}}-1 \right)}. \nonumber
\end{eqnarray}
Here, $\Delta E$ defines the energy interval between the resonance energy $E_R$ and the energies where the Maxwell-Boltzmann function crossing the Breit-Wigner
function. It is obvious that the dominate variation takes place in the third exponential term. Given a condition of $\Delta E/E_R$ = 5 \%, for
$p(p,\gamma)$ reaction, $\delta_{\Gamma_a\Gamma_\gamma}$ are only  about 8\%, 5\%, 4\% for resonances at $E_R$=0.1, 0.5, 1.0 MeV, respectively; while for
$^{12}C(p,\gamma)$ reaction, $\delta_{\Gamma_a\Gamma_\gamma}$ are respectively increased to 58\%, 24\%, 18\% for the above resonance energies. Clearly the variation
$\delta_{\Gamma_a\Gamma_\gamma}$ is very sensitive to the projectile-target combination via quantity $c_2$. Therefore, the THIRD assumption is appropriate only for
certain projectile-target combinations for a resonance with $\Gamma/E_R$$<$10 \%.

Actually, the {\tt RRR} per particle pair for a narrow resonance can be directly written as (use Eqs.~\ref{eq:2}~\&~\ref{eq:3})
\begin{eqnarray}
\langle \sigma v \rangle = && \sqrt{2\pi}\frac{\hbar^2\omega}{(\mu kT)^{3/2}} \nonumber \\
&& \times \int^{\infty}_{0} \mathrm{exp} \left(-\frac{E}{kT}\right) \frac{\Gamma_a \Gamma_b}{(E-E_R)^2+(\Gamma/2)^2} dE.
\label{eq:7}
\end{eqnarray}
If one assumes quantity $\mathrm{exp}(-E/kT)$ as a constant over the resonance region, Equ.~\ref{eq:7} is equivalent to Equ.~\ref{eq:6}. But, obviously this is not a
good assumption for certain circumstances (see discussion for the FIRST assumption). As a conclusion, except the SECOND reasonable assumption, the FIRST and THIRD
assumptions only hold for certain circumstances.

Now, let's turn to Equ.~\ref{eq:5}. Here an approximation ({\em {i.e.}}, $\arctan(\frac{E_R}{\Gamma/2})$$\simeq$$\frac{\pi}{2}$) was utilized, and hence one can get
\begin{eqnarray}
\int^{\infty}_{0} \frac{1}{(E-E_R)^2+(\Gamma/2)^2} dE \simeq \frac{\pi}{\Gamma/2}. \nonumber
\end{eqnarray}
This approximation causes about a 3\% error for a condition of $\Gamma/E_R$= 10\%, and thus it can be regarded as a reasonable one. The integration ranges, from 0 to
$\infty$, were used in Equ.~\ref{eq:5}, and therefore the quantity $\Gamma_a\Gamma_b$ can't be considered as a constant and taken outside the integral. The quantity
$\Gamma_a\Gamma_b$ can be roughly considered as a constant over a very narrow energy interval, and therefore the integration ranges on Equ.~\ref{eq:5} should be
physically restricted~\cite{bib:ang99}. A general equation has been derived here,
\begin{eqnarray}
\int^{E_R+n\frac{\Gamma}{2}}_{E_R-n\frac{\Gamma}{2}} \frac{1}{(E-E_R)^2+(\Gamma/2)^2} dE = \frac{2}{\Gamma/2} \mathrm{arctan}(n).
\label{eq:8}
\end{eqnarray}
Ratios between the values of quantity $\frac{\pi}{\Gamma/2}$ and those of quantity $\frac{2}{\Gamma/2}\mathrm{arctan}(n)$ are 0.5, 0.7, 0.8 for $n$=1,2,4, respectively.
This effect can be seen in the following Table~\ref{table1} as well. Thus for a small value of $n$, the `look reliable' derivation of Equ.~\ref{eq:5} is
inappropriate in the integration-range point of view; while for a large $n$, although the difference becomes small, the above-mentioned assumptions might be
inappropriate in the `narrow' point of view. Therefore, the classical analytic Eqs.~\ref{eq:4}-\ref{eq:6} are suggested not to be utilized in the future
{\tt RRR} calculations without caution.

\section{Numerical Integration}
In the past, some researchers utilized the numerical integration of Equ.~\ref{eq:1} with a broad-resonance formula to calculate the
resonant-reaction-rate ({\tt RRR}) of a narrow resonance, while others still utilized the simple analytic expression of Equ.~\ref{eq:6}.
In order to check their differences, as an example, the {\tt RRR} of a key stellar reaction
$^{21}$Na(p,$\gamma$)$^{22}$Mg is discussed below. There is a known resonant state in the compound nucleus $^{22}$Mg at $E_R$=0.821 MeV
($J^{\pi}$=$1^{+}$, $\ell$=0) with a proton width of $\Gamma_p$=16 keV and a resonant strength of $\omega \gamma$=0.556 eV
(here $\Gamma_\gamma$$\approx$$\frac{\omega\gamma}{\omega}$=1.48 eV, and hence $\Gamma$$\approx$$\Gamma_p$)~\cite{bib:dau04}. The corresponding
$\gamma$-transition to the ground state in $^{22}$Mg is a pure $M$1 transition ({\em i.e.}, 1$^{+}$$\rightarrow$0$^{+}$) with a multipolarity $L$=1.
According to the previous definition, this resonance ($\Gamma/E_R$$\simeq$2\%) can be `reasonably' treated as a narrow one.

Figure 1 shows the amplitudes of relevant functions used in the equations for this resonance state at the `Gamow-peak' temperature $T$=1.6 GK. The narrow-resonance
cross section $\sigma_{BW}^{N}(E)$ is calculated by assuming the constant partial widths $\Gamma_p$, $\Gamma_\gamma$ (and $\Gamma$). Obviously the tendencies for this
cross section and the corresponding integrand curve (indicated by a grey thick solid line) are physically unreal towards lower energies. The reason is the assumption
of constant widths is invalid out of the resonance region. If a broad-resonance formula is used in the calculation, the integrand curve (indicated by a thick solid
line) become physically reasonable at low energies.
Here, the broad-resonance cross section $\sigma_{BW}^B(E)$ can be written as, ({\em i.e.}, Equ.~(4.59))
in~\cite{bib:rol88},
\begin{equation}
\sigma(E)_{BW}^B = \sigma_R\frac{E_R}{E}\frac{\Gamma_p(E)}{\Gamma_p(E_R)}\frac{\Gamma_\gamma(E)}{\Gamma_\gamma(E_R)} \frac{(\Gamma_R/2)^2}{(E-E_R)^2+[\Gamma(E)/2]^2}.
\label{eq:9}
\end{equation}
The energy dependence of the proton and $\gamma$-ray partial widths are given by
\begin{eqnarray}
\Gamma_p(E)=\Gamma_p(E_R)\frac{P_{\ell=0}(E)}{P_{\ell=0}(E_R)}, \nonumber
\end{eqnarray}
\begin{eqnarray}
\Gamma_\gamma(E)=\Gamma_\gamma(E_R)B_\gamma\left(\frac{E+Q_{p\gamma}-E_{xf}}{E_R+Q_{p\gamma}-E_{xf}} \right)^{2L+1}, \nonumber
\end{eqnarray}
where $Q_{p\gamma}$ is the (p, $\gamma$) reaction $Q$-value ($Q$=5.508 MeV~\cite{bib:dau04}), $B_{\gamma}$ is the primary $\gamma$-ray branching ratio to the final state at
$E_{xf}$ (Here, $B_{\gamma}$ is assumed to be unity and $E_{xf}$=0 for the ground state). The penetration factors, $P_\ell$, are calculated by using a
{\tt RCWFN}~\cite{bib:bar74} code with a channel-radius parameter of $a_0$=1.35 fm.

The numerically and analytically calculated resonant-reaction-rates ({\tt RRR}s) are compared in Table 1, where the corresponding
ratios were listed for different temperatures and integration ranges used in Equ.~\ref{eq:7}. It can be seen that two methods give almost the same results at
`Gamow-peak' temperature $T$=1.6 GK for a full-range integration (0$\rightarrow$$\infty$). For narrow integration ranges, the analytic results are larger than the
numerical ones. This shows that the integration should be computed for a full range, and it is not a difficult task anymore with current computers. Notably, at
lower temperatures, the numerical results become much larger than the analytic ones (see the 2nd column in Table 1). This implies a higher-energy resonance can
contribute to the reaction rate at low temperature significantly (or tremendously) more than the previous analytical result. Furthermore, presume this $E_R$=0.821 MeV
resonance locating at 0.160, 0.323 MeV, the calculated $Ratio$s for a full-range integration are 0.78, 0.90, respectively, at the relevant `Gamow-peak' temperatures
of 0.14, 0.4 GK. This conclusion demonstrates that the previous analytic results may be underestimated by a considerable amount for a resonance with a several keV
width even at their `Gamow-peak' temperatures. Although the condition $\Gamma/E_R$$<$10 \% still holds in these cases, $\delta_{MB}$ varies from 6 \% to 37 \% for
the temperatures listed in Table 1. It indicates that, in the present case, the resonance width is not narrow enough to use the analytic equations at low temperatures,
and hence another restriction on the width should be set, {\em e.g.}, $\Gamma$$\leq$ 3.4 keV for a temperature of 0.2 GK, and it is about five times narrower than
the experimental value ($\Gamma$=16 keV).

\section{Conclusions}
The validity of the resonant-reaction-rate ({\tt RRR}) equations for an isolated narrow resonance have been reexamined, and it reminds us not to use those analytic
equations without careful examination. As a suggestion, it's better to use the broad-resonance equation to calculate the {\tt RRR} numerically even for a narrow
resonance. In addition, the {\tt RRR} is usually calculated by using the measured $E_R$ and $\omega\gamma$ analytically in the popular direct measurement approach.
We think this kind of calculation is inaccurate at low temperatures for a resonance of a few keV width. Because there are still researchers using the analytic
equations without rigorous examinations so far, it is worthwhile to address this issue clearly and it will be helpful for the communities. Two years ago, a new
book ``{\it Nuclear Physics of Stars\/}"~\cite{bib:iliadis} was published, where the definition of a narrow-resonance seems reasonable. But, we should realize that
a question is hidden in this sort of calculation, {\em i.e.}, is it physically appropriate to extend the broad-resonance Breit-Wigner formulism ({\em esp.} for a
narrow resonance) down to very low energy? since the energy gap between the resonance and the Gamow-peak is quite large, it's more than hundreds or thousands of
resonance width sometimes. However, we can't address this question in the present work.
In addition, some authors~\cite{bib:ang99,bib:iliadis} thought that the product of Maxwell-Boltzmann distribution and cross section could
give rise to another maximum caused by the low-energy wing of a resonance but, our calculation shows the appearance of this low-energy maximum depends on the
temperature condition, for instance, there is no such maximum (but a plateau) in our example beyond $T$$\sim$0.5 GK (of course, no maximum at
$T$=1.6 GK in Fig.~\ref{fig1}).

\begin{acknowledgement}
This work is financially supported by the ``100 Persons Project" of Chinese Academy of Sciences, and by the Major State Basic Research Development Program of China
(2007CB815000).
\end{acknowledgement}

\newpage
\begin{table}[t]
\caption{\label{table1} Ratios of the calculated resonant-reaction-rates ({\tt RRR}s) using the numerical integration of Equ.~\ref{eq:1} with a broad-resonance formula
and using the analytic Equ.~\ref{eq:6}. The ratios are listed for different temperatures and integration ranges, respectively. Additionally the values of $\delta_{MB}$
are listed for different temperatures as well.}
\begin{center}
\begin{tabular}{cccccc}
\hline \hline
\multicolumn{1}{c} {} &\multicolumn{4}{c}{$Ratio=\mathrm{Numerical}/\mathrm{Analytic}$} &\multicolumn{1}{c}{}\\
\cline{2-5}
$T_9$(GK) & 0 $\rightarrow$ $\infty$ & $E_R \pm \frac{\Gamma}{2}$ & $E_R \pm 2\frac{\Gamma}{2}$ & $E_R \pm 4\frac{\Gamma}{2}$ & $\delta_{MB}(\%)$\\
\hline
0.2 & 2.4$\times 10^{7}$ & 0.52 & 0.76 & 1.02 & 37  \\
0.4 & 6.6                & 0.50 & 0.72 & 0.87 & 21  \\
0.5 & 1.6                & 0.50 & 0.71 & 0.86 & 17  \\
1.6 & 1.0                & 0.50 & 0.71 & 0.84 & 5.6 \\
\hline \hline
\end{tabular}
\end{center}
\end{table}


\begin{thebibliography}{}
\bibitem{bib:fow67}
W.A. Fowler, G.R. Caughlan, and B.A. Zimmerman, Ann. Rev. Astron. Astrophys. \textbf{5}, 525 (1967).
\bibitem{bib:cla83}
D.D. Clayton, \textit{Principles of Stellar Evolution and Nucleosynthesis}, (University of Chicago Press, Chicago, 1983)
\bibitem{bib:rol88}
C.E. Rolfs and W.S. Rodney, \textit{Cauldrons in the Cosmos}, (University of Chicago Press, Chicago, 1988)
\bibitem{bib:bar74}
A.R. Barnett {\it et al.}, Comput. Phys. Commun. {\bf 8}, 377 (1974).
\bibitem{bib:ang99}
C. Angulo {\it et al.}, Nucl. Phys. {\bf A656}, 3 (1999).
\bibitem{bib:dau04}
J.M. D'Auria, {\it et al.}, Phys. Rev. C {\bf 69}, 065803 (2004).
\bibitem{bib:iliadis}
C. Iliadis, \textit{Nuclear Physics of Stars}, (Wiley-vch, Germany, 2007)
\end{thebibliography}
\end{document}